%% file: sample701.tex
\documentclass[twocolumn, trackchanges]{aastex701}
\usepackage{amsmath}
\usepackage{multirow}

\begin{document}

\title{Origin of the stellar Fe K$\alpha$ line clarified with FUV and X-ray observations \\ of a superflare on the RS Canum Venaticorum–type Star UX Arietis}

\author[orcid=0000-0003-3085-304X]{Shun Inoue}
\affiliation{Department of Physics, Kyoto University, Kitashirakawa-Oiwake-cho, Sakyo-ku, Kyoto, 606-8502, Japan}
\email[show]{inoue@cr.scphys.kyoto-u.ac.jp}  

\author[orcid=0000-0002-0207-9010]{Wataru Buz Iwakiri} 
\affiliation{International Center for Hadron Astrophysics, Chiba University, Inage-ku, Chiba, 263-8522 Japan}
\email{iwakiri@chiba-u.jp}

\author[orcid=0000-0002-0636-4687]{Tomoki Kimura} 
\affiliation{Department of Physics, Faculty of Science, Tokyo University of Science, Tokyo 162-8601, Japan}
\email{kimura@rs.tus.ac.jp}

\author[orcid=0000-0003-1244-3100]{Teruaki Enoto} 
\affiliation{Department of Physics, Kyoto University, Kitashirakawa-Oiwake-cho, Sakyo-ku, Kyoto, 606-8502, Japan}
\affiliation{RIKEN Center for Advanced Photonics (RAP), 2-1 Hirosawa, Wako, Saitama 351-0198, Japan}
\email{enoto.teruaki.2w@kyoto-u.ac.jp}

\author[orcid=0000-0002-0412-0849]{Yuta Notsu} 
\affiliation{Laboratory for Atmospheric and Space Physics, University of Colorado Boulder, 3665 Discovery Drive, Boulder, CO 80303, USA}
\affiliation{National Solar Observatory, 3665 Discovery Drive, Boulder, CO 80303, USA}
\email{yuta.notsu@colorado.edu}

\author[orcid=0000-0003-1518-2188]{Hiroyuki Uchida} 
\affiliation{Department of Physics, Kyoto University, Kitashirakawa-Oiwake-cho, Sakyo-ku, Kyoto, 606-8502, Japan}
\email{uchida@cr.scphys.kyoto-u.ac.jp}

\author[orcid=0000-0001-7515-2779]{Kenji Hamaguchi} 
\affiliation{CRESST II and X-ray Astrophysics Laboratory, NASA’s Goddard Space Flight Center, Greenbelt, MD 20771, USA}
\affiliation{Department of Physics, University of Maryland, Baltimore County, Baltimore, MD, USA}
\email{kenjih@umbc.edu}

\author[orcid=0000-0002-1276-2403]{Shin Toriumi} 
\affiliation{Institute of Space and Astronautical Science, Japan Aerospace Exploration Agency, Yoshinodai, Chuo-ku, Sagamihara, Kanagawa 252-5210, Japan}
\email{toriumi.shin@jaxa.jp}

\author[orcid=0000-0001-6468-6812]{Atsushi Yamazaki} 
\affiliation{Institute of Space and Astronautical Science, Japan Aerospace Exploration Agency, Yoshinodai, Chuo-ku, Sagamihara, Kanagawa 252-5210, Japan}
\email{yamazaki@stp.isas.jaxa.jp}

\author[orcid=0000-0003-3386-6794]{Fuminori Tsuchiya} 
\affiliation{Planetary Plasma and Atmospheric Research Center, Graduate School of Science, Tohoku University, Sendai, Japan}
\email{tsuchiya.f@tohoku.ac.jp}

\author[orcid=0000-0002-2759-7682]{Go Murakami} 
\affiliation{Institute of Space and Astronautical Science, Japan Aerospace Exploration Agency, Yoshinodai, Chuo-ku, Sagamihara, Kanagawa 252-5210, Japan}
\email{go@stp.isas.jaxa.jp}

\author[orcid=0000-0001-5451-9367]{Kazuo Yoshioka} 
\affiliation{Department of Complexity Science and Engineering, Graduate School of Frontier Science, The University of Tokyo, Kashiwa, Japan}
\email{kazuo.yoshioka@edu.k.u-tokyo.ac.jp}

\author[orcid=0009-0008-6187-8753]{Zaven Arzoumanian} 
\affiliation{Astrophysics Science Division, NASA’s Goddard Space Flight Center, Greenbelt, MD 20771, USA}
\email{zaven.arzoumanian-1@nasa.gov}

\author[orcid=0000-0001-7115-2819]{Keith Gendreau} 
\affiliation{Astrophysics Science Division, NASA’s Goddard Space Flight Center, Greenbelt, MD 20771, USA}
\email{keith.c.gendreau@nasa.gov}

% \collaboration{all}{The Terra Mater collaboration}

\begin{abstract} 
Fluorescence line diagnostics of the Fe K$\alpha$ line at $\sim 6.4$ keV observed in both solar and stellar flares can constrain the latitude and size of the flare loop, even in the absence of imaging observations. 
However, they are hampered by the unresolved origin of stellar Fe K$\alpha$ lines: i.e., it is unclear which of the two mechanisms—photoionization by hard X-ray photons or collisional ionization by non-thermal electrons—is the dominant process. 
We present clear evidence for the photoionization origin based on simultaneous far ultraviolet (FUV) and soft X-ray observations of a superflare on the RS Canum Venaticorum–type Star UX Arietis with Extreme ultraviolet spetrosCope for ExosphEric Dynamic (EXCEED; 900$-$1480 \AA) onboard Hisaki and Neutron Star Interior Composition Explorer (NICER; 0.2$-$12 keV). 
The flare started at 22:50 UT on 2018 November 15 and released $2 \times 10^{36}$ erg in the 900$-$1480 \AA $\,$ band and $3 \times 10^{36}$ erg in the 0.3$-$4 keV band.
The FUV emission, a proxy for non-thermal activity, peaked approximately 1.4 hours before the soft X-rays. 
In contrast, the Fe K$\alpha$ line, detected at a statistical significance of $5.3 \sigma$ with an equivalent width of $67^{+28}_{-20}$ eV, peaked simultaneously with the thermal X-ray maximum rather than the non-thermal FUV peak—strongly supporting the photoionization hypothesis.
Radiative transfer calculations, combined with the observed Fe K$\alpha$ line intensity, further support the photoionization scenario and demonstrate the potential of this line to provide the flare geometry.
\end{abstract}



\section{Introduction} 
The impact of stellar magnetic activity on the surrounding environment, known as space weather, has been explored both in our solar and in exoplanetary systems \citep[e.g.][]{Airapetian_2020}.
This activity is a key factor for planetary habitability, as planets are exposed to high-energy radiation and energetic particles from their host star.
The most significant difference in the study of solar and stellar flares is the availability of the spatially resolved observations.
We know that the impact of a solar flare on Earth's environment strongly depends on the position and direction of its flare loop relative to the Earth. 
For example, when Coronal Mass Ejections (CMEs) associated with solar flares erupt from the disk plane directly heading to the Earth, large magnetic storms cause ionospheric disturbances \citep{Boteler_2019}.
On the other hand, the quantitative evaluations of stellar flare effects on exoplanet environments are limited since the size and stellar flaring regions are not determined and eruption directions cannot be constrained, because of the lack of spatially-resolved informations except for rare cases of eclipsing binaries \citep{Schmitt_2003, Nakayama_2025}.

Fe fluorescence lines from neutral to mildly-ionized ions in X-rays, such as Fe K$\alpha$ at $6.4-6.6$ keV, have been used to investigate the geometry and motion of surrounding matters around a variety of objects \citep[e.g., ][]{Hamaguchi_2005, Watanabe_2006, Ishida_2009, Hayashi_2011, Odaka_2016, Hayashi_2018, Hitomi_2018, Kawamuro_2019, Hayashi_2021, Noda_2023, Mochizuki_2024, XRISM_2024}.
The Fe K$\alpha$ lines are emitted when X-ray photons or electrons with energies above the Fe K-edge (7.11 keV) are irradiated and stimulate ions, producing X-ray photons with energies at transitions from upper levels to K-shell orbits.

The Fe K$\alpha$ line has been observed in solar and stellar flares and is considered a result of the illumination of the stellar surface via X-ray photons or non-thermal electrons by the flare loop \citep{Neupert_1967, Neupert_1971, Doschek_1971, Bai_1979, Feldman_1980, Culhane_1981, Parmar_1982, Tanaka_1982, Tanaka_1983, Tanaka_1984, Parmar_1984,Tanaka_1985, Emslie_1986, Zarro_1992, Phillips_1995, Osten_2007, Testa_2008, Osten_2010, Huenemoerder_2010, Karmakar_2017, Kurihara_2025, Inoue_2025, Suzuki_2025}. 
If the origin of the Fe K$\alpha$ line is the inner-shell ionization of the neutral or low-ionized iron ions in the photosphere by X-ray photons, the Fe K$\alpha$ line intensity and profile of solar and stellar flares can be used to estimate the latitude and size of the flare loop, supported by radiative transfer calculations \citep{Kowalski_2024} because the fluorescence efficiency has a dependency on the loop height and inclination angle between the line of sight and flare loop \citep{Testa_2008}. 
On the other hand, if  non-thermal electrons are the primary origin of the Fe K$\alpha$ line, the Fe K$\alpha$ spectroscopy could be used as a discriminator between the non-thermal and thermal models \citep{Emslie_1986}.

The latitude of stellar flares is also key to understanding the generation mechanism of the stellar magnetic fields.
In the solar case, the distribution of sunspots and flares is concentrated at low latitudes \citep{Abdel-Sattar_2018}, a phenomenon explained by a dynamo model \citep{Nandy_2002}.
However, there are still few examples of the Fe K$\alpha$ usage to investigate stellar flare geometry \citep{Osten_2007, Testa_2008, Osten_2010, Huenemoerder_2010, Karmakar_2017} because there remain two possible emission mechanisms of the Fe K$\alpha$ line in solar and stellar flares: photoionization by hard X-ray photons from the flare loop and collisional ionization by nonthermal electrons.
It remains unknown which of these emission mechanisms is dominant.
\citet{Osten_2010} discussed the possibility that the Fe K$\alpha$ line observed during a stellar flare on the M-dwarf EV Lac could be produced not only by fluorescence but also by collisional excitation. However, \citet{Osten_2016} also pointed out that this detection may have been a calibration artifact caused by charge trapping in Swift \citep{Pagani_2011}.

RS Canum Venaticorum (RS CVn) type stars are well-known magnetically active binaries, which frequently produce large superflares \citep[e.g.,][]{Patkos_1981, Rodono_1986, Rodono_1987, Walter_1987, Tsuru_1989, Doyle_1991, Mathioudakis_1992, Kuerster_1996, Endl_1997, Gudel_1999, Osten_1999,Osten_2000, Franciosini_2001, Gudel_2002, Osten_2003, Osten_2004, Brown_2006, Osten_2007, Pandey_2012, Tsuboi_2016, Cao_2017, Cao_2019, Cao_2020, Sasaki_2021, Kawai_2022, Inoue_2023, Karmakar_2023, Kurihara_2024, Inoue_2024b, Gunther_2024, Cao_2024a, Cao_2024b, Didel_2025, Cao_2025, Xu_2025}. 
In this article, we report the far ultraviolet (FUV) and soft X-ray observation of a superflare from UX Arietis (UX Ari), one of the most prominent RS CVn type stars.
We describe observation and data reduction (Section \ref{sec:obsrvation_reduction}), analysis and results (Section \ref{sec:analysis_result}), and discussion and conclusion (Section \ref{sec:discussion_conclusion}). 
In this work, the error ranges indicate 90\% confidence level unless otherwise indicated.

\begin{figure*}[]
 \begin{center}
  \plotone{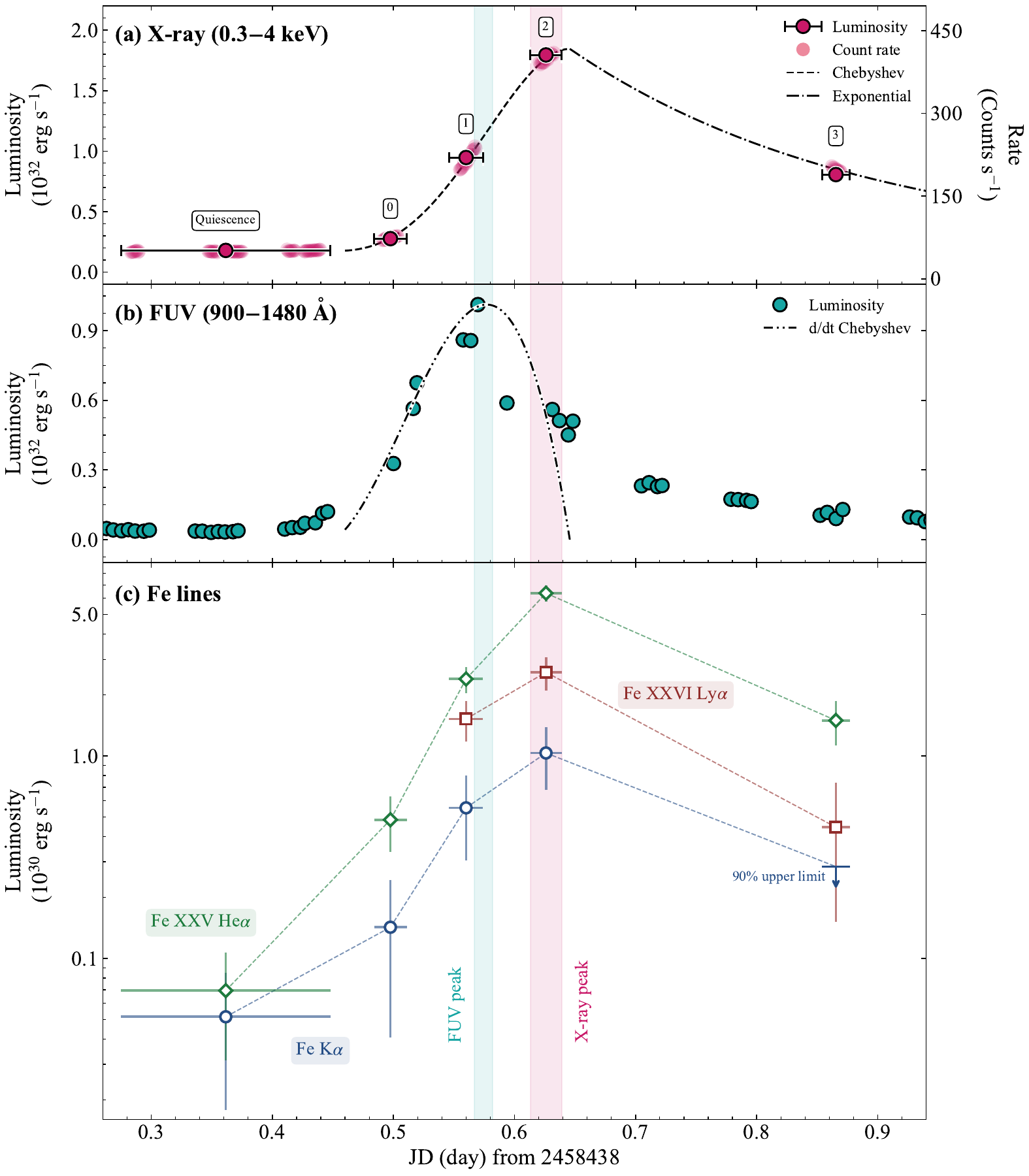}
  \caption{Light curves of the superflare of UX Ari on 16 November 2018. (a) The $0.3-4$ keV NICER count rate (cps; $\mathrm{counts \: s^{-1}}$) binned at 64 sec of UX Ari and $0.3-4$ keV luminosity binned at $\sim90$ min ISS orbit. The time origin of JD 2458438 corresponds to 12:00 UT 2018 November 15. The one standard deviation statistical uncertainties are smaller than the symbol size. The black dashed and dash-dot lines indicate the 4th order Chebyshev polynomial and exponential function fitted to the impulsive and decay phases of the count rate light curve, respectively. Interval numbers in our spectral analysis are also shown. (b) Background-subtracted $900-1480$ $\mathrm{\AA}$ luminosity of UX Ari measured by Hisaki. Black dash-dot-dot line shows the time derivative of the 4th order Chebyshev polynomial in panel (a). The derivative curve of the Chebyshev function is normalized to the FUV peak luminosity. (c) Time variation of the Fe line luminosity in X-rays. The blue circles, green diamonds and red squares are the luminosity of the Fe K$\alpha$, Fe XXV He$\alpha$, and Fe XXVI Ly$\alpha$ lines, respectively. The red and blue shaded area show the X-ray and FUV flare peaks, respectively. The down arrow of the Fe K$\alpha$ line means the 90\% upper limit and the errorbars indicate 90\% confidence level.
  \label{Figure1}}
 \end{center}
\end{figure*}

\section{Observation and Data Reduction} \label{sec:obsrvation_reduction}
UX Ari is located at 50.5 pc \citep{Bozzo_2024} and consists of a K0 subgiant primary component (Aa), a G5 main-sequence companion (Ab), and a G5V third star (B) \citep{Hummel_2017}.
The rotation axis of UX Ari Aa is inclined by $i_{r} = 59.2^\circ$ with respect to our line of sight \citep{Duemmler_2001}.
Large superflares on UX Ari have been observed at various wavelengths \citep{Tsuru_1989, Cao_2017, Kurihara_2024, Cao_2025}.

For this study, we simultaneously performed FUV (900$-$1480 $\mathrm{\AA}$) and soft X-ray (0.2–12 keV) observations of a superflare on UX Ari (Figure \ref{Figure1}), starting at 22:50 UT on 2018 November 15 and releasing $2 \times 10^{36}$ erg in the 900$-$1480 \AA $\,$ band and $3 \times 10^{36}$ erg in the 0.3$-$4 keV band.
We used the UV spectrometer Extreme ultraviolet spetrosCope for ExosphEric Dynamics \citep[EXCEED; ][]{Yoshioka_2013} on board JAXA's Hisaki satellite and NASA’s Neutron Star Interior Composition Explorer \citep[NICER; ][]{2016SPIE.9905E..1HG}.

\subsection{NICER}
The NICER observation of UX Ari was conducted from 18:43 UT on 15 November 2018  to 08:51 on 16 November 2018.
We downloaded the NICER data from the HEASARC archive (ObsID: 1100380107 and 1100380108).
The ObsID 1100380107 and 1100380108 data covered the quiescence and flare phase, respectively.
The data were processed and calibrated in the standard manner, same as that in \cite{Inoue_2024b, Inoue_2025}, with \texttt{nicerl2} in HEASoft ver. 6.32.1 \citep{heasoft_2014} and the calibration database (\texttt{CALDB}) version \texttt{xti20240206}.
We used two filtering criteria for \texttt{nicerl2}: (a) overshoot count rate range of 0$-$5; (b) cut off rigidity higher than 1.5 $\mathrm{GeV \: c^{-1}}.$
After the good time interval (GTI) filtering, we extracted light curves from the filtered event files with \texttt{xselect}.
Then, we generated ObsID-averaged source and background spectra of ObsID 1100380107 with \texttt{nicerl3-spect} and the 3C50 model \citep{Remillard_2022} and time-resolved spectra at each GTI of ObsID 1100380108 with \texttt{nimaketime}, \texttt{niextract-event} and \texttt{nicerl3-spect}.
We numbered GTIs of ObsID 1100380108 as Interval 0, 1, 2, and 3 by time (Figure \ref{Figure1}a).

\subsection{Hisaki}
Hisaki is the FUV space telescope and originally a satellite for planetary science in our solar system \citep[e.g.,][]{Tsuchiya_2015, Murakami_2016, Kimura_2018}, but it can observe stellar objects within the ecliptic plane of $\pm 10^{\circ}$. EXCEED \citep{Yoshioka_2013} on board Hisaki covers 520$-$1480 \AA \hspace{0.5pt} with the 3$-$5 \AA \hspace{0.5pt} Full Width at Half Maximum (FWHM).

The Hisaki FUV imaging spectral data is a Level-2 product derived from raw observation data through a processing pipeline developed by \cite{Kimura_2019}. The light curve data used in this study was processed from this imaging spectral data, as outlined in Section 6.1 of \cite{Kimura_2019}.

\section{Analysis and Result} \label{sec:analysis_result}
\subsection{Light Curve}\label{subsec:analysis_result_lightcurve}
We fit the impulsive phase of the X-ray light curve ($0.4 <$ day $< 0.7$ in Figure \ref{Figure1}a) with the 4th-order Chebyshev polynomial. The fitted function is expressed as 
\begin{equation}
\begin{split}
C (t) \sim &\; 183 + 191 T_{1} (t) + 57 T_{2} (t) \\
           & - 12 T_{3} (t) - 11 T_{4} (t), \;\; (t_{\mathrm{start}} < t < t_{\mathrm{peak}})
\end{split}
\end{equation}
where $C (t)$ is the 0.3$-$4 keV count rate, $t$ is the day from JD 2458438, $T_{n} (t)$ is 
\begin{equation}
\begin{split}
T_{n} (t) = \frac{(t+\sqrt{t^{2}-1})^{n} + (t-\sqrt{t^{2}-1})^{n}}{2}
\end{split}
\end{equation}
and $t_{\mathrm{start}}$ and $t_{\mathrm{peak}}$ are the X-ray flare start ($t_{\mathrm{start}} \sim 0.45$ day) and peak time ($t_{\mathrm{peak}} \sim 0.65$ day), respectively.
We also fit the decay phase of the X-ray light curve ($t > 0.7$ in Figure \ref{Figure1}a) with the sum of the exponential function and quiescence component as
\begin{equation}
    C (t) = C_{q} + C_{p} \exp(-t/\tau_{d}) \; \; \; \; \; \; \; \;  (t_{\mathrm{peak}}<t),
\end{equation}
where $C_{q}$ and $C_{p}$ are the median count rate during the quiescence phase and the flare peak count rate.
The $e$-folding time $\tau_{d}$ of the X-ray decay phase is derived to be $\sim 0.24$ day from the fitting.

To confirm the existence of Neupert effect (see Section \ref{sec:discussion_conclusion} for details), we overplot the time derivative of the Chebyshev polynomial fitted to the impulsive phase of the X-ray light curve in the FUV light curve (Figure \ref{Figure1}b). 
The norm of the derivative curve is scaled by the ratio of the peak FUV luminosity to the peak value of the derivative of the Chebyshev polynomial.
The FUV emission in Figure \ref{Figure1}b peaked $\sim 1.4$ hours earlier than the soft X-ray flare peak, and the time derivative of the soft X-ray emission during the impulsive phase well correlated with the FUV light curve. 
We confirm that the light curves of each FUV line and continuum in 900$-$1480 \AA $\,$ is also correlated with the time derivative of the X-ray light curve (Figure \ref{FigureE0}).

\begin{figure*}
\begin{center}
\includegraphics[width=16cm]{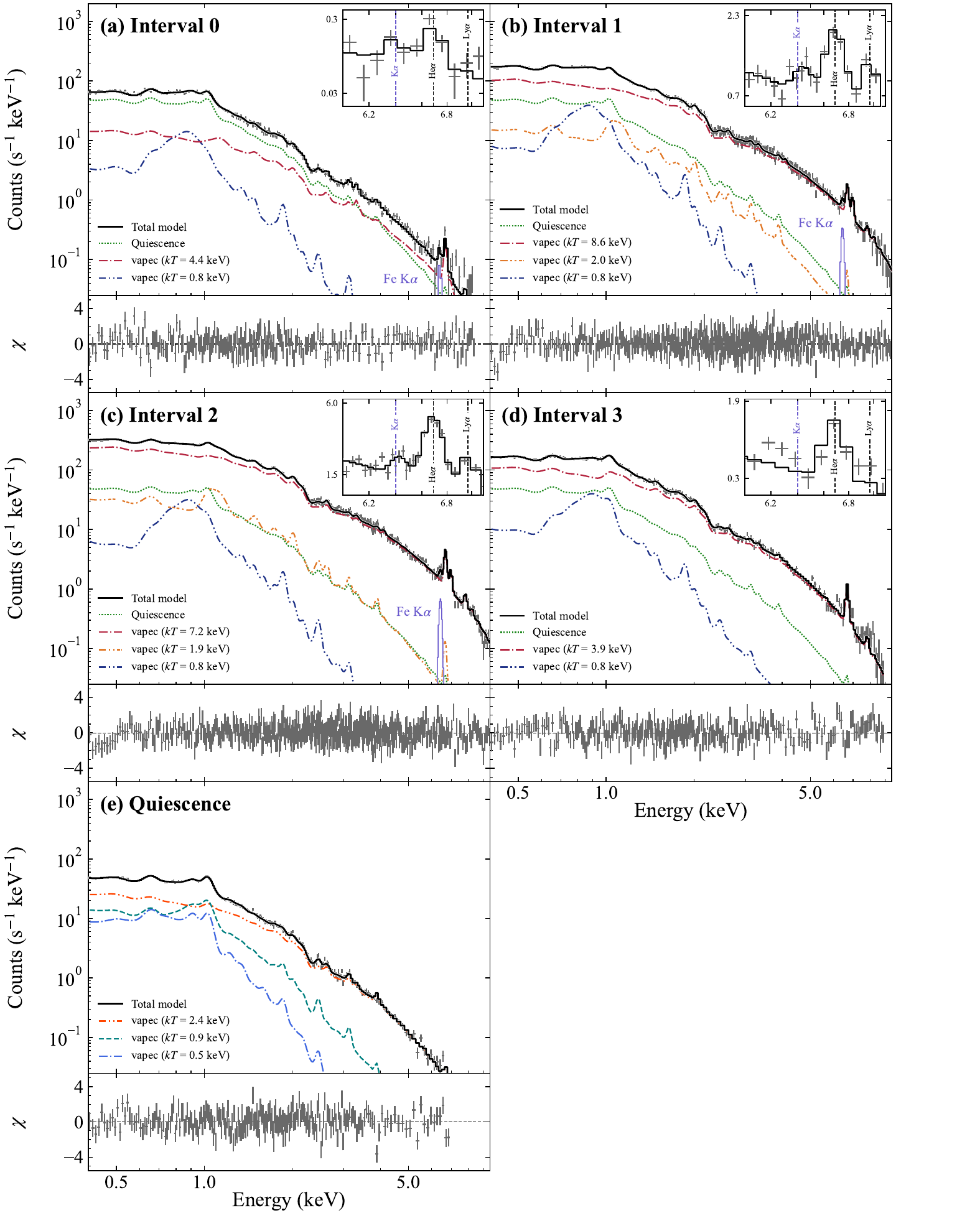}
\end{center}
\caption{
Background-subtracted and response-uncorrected X-ray spectra at each interval shown in Figure \ref{Figure1}a. 
  Panels (a-d) are spectra during the flare, whereas panel (e) is the spectrum during the quiescence. (a-d) The best-fit curves for two-temperature \texttt{vapec} models with the fixed quiescence component are shown by solid black lines. Green dotted, red dash-dot, blue dash-dot-dot, lines represent the quiescence, high temperature, and low temperature flare component, respectively. The right upper inset panels
show the enlarged spectrum around the Fe K$\alpha$ at 6.4 keV, Fe XXV He$\alpha$ at 6.7 keV, and Fe XXVI Ly$\alpha$ at 6.9 keV. (e) Orange dash-dot-dot, green dashed, and blue dash-dot lines represent the high-, medium-, and low-temperature components, respectively.
}
\label{FigureE1}
\end{figure*}

\begin{figure}
\begin{center}
\includegraphics[width=8.5cm]{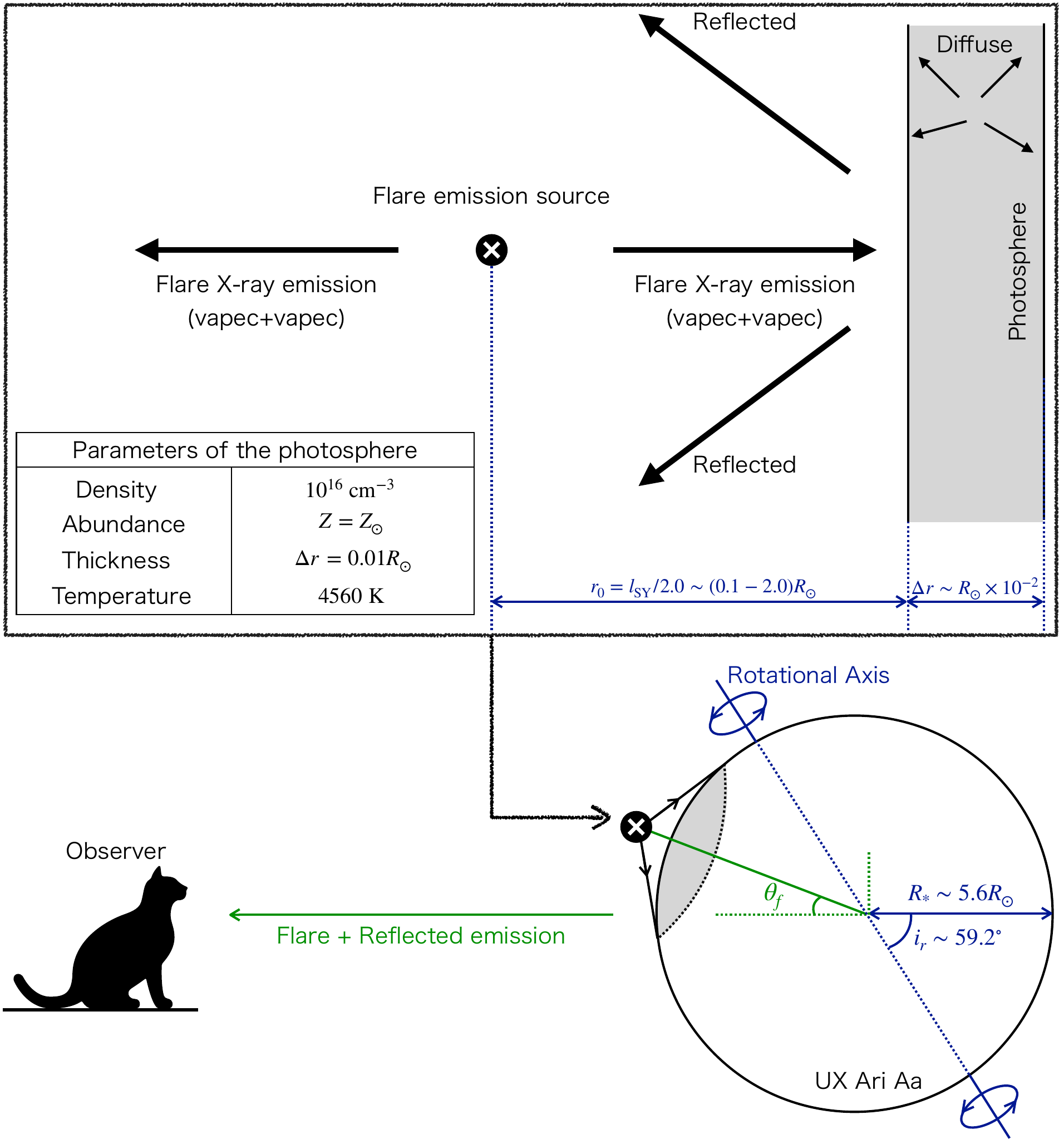}
\end{center}
\caption{
Schematic diagram of the setup of the \texttt{SKIRT} simulation.
}
\label{Figure3}
\end{figure}

\subsection{X-ray spectrum}\label{subsec:analysis_result_spectrum}
Figure \ref{FigureE1} shows all NICER X-ray spectra of each interval during the flare and quiescence shown in the light curve of Figure \ref{Figure1}.
First, we fit the 0.4$-$6.9 keV quiescent spectrum (ObsID 1100380107) with three-temperature collisionally ionized equilibrium (CIE) components (\texttt{vapec}) with interstellar absorption (\texttt{tbabs}). 
Single- and two-temperature CIE models yielded statistically unacceptable fits. 
We set the abundances of C, Ne, Mg, Si, S, and Fe free, whereas those of N and O equal to that of C because their first ionization potential (FIP) is close.
We fixed the abundances of He, Al, Ar, Ca, and Ni to the solar value \citep{Anders_1989}.
We set the abundances of each element equal among the three temperature components.
The best-fit temperatures of the three components of the quiescence were 28.11 MK, 10.54 MK, and 5.86 MK (Table \ref{TableE1}).

Based on the results of the spectral analysis of the quiescence spectrum, we analyzed the spectra for each interval during the flare (ObsID 1100380108; Interval 0–3).
We fixed the best-fitting model of the quiescent component and fitted the flare component with two- or three-temperature CIE models (\texttt{vapec}).
For Intervals 1 and 2, three-temperature CIE models were required because two-temperature fits left strong residuals around the Fe L$\alpha$ line at 1.0–1.2 keV (Appendix \ref{app:FeLa}). 
This medium-temperature component ($kT \sim 2.0$ keV) is necessary to reproduce the Fe L$\alpha$ line intensity. This result is consistent with hydrodynamic flare loop simulations \citep{Hamaguchi_2023}, which suggest that cool flare plasmas contain multiple temperature components.
We set the abundances of each element equal among the flare components. 

The best-fit temperatures of the three components for the flare plasma were 100.17, 22.85, and 9.05 MK at the FUV flare peak (Interval1) and 83.64, 22.27, and 9.28 MK at the X-ray flare peak (Interval 2).
The emission of the hottest component dominates the spectrum above 2 keV and its temperature of 83.64 MK at Interval 2 is consistent with the existence of the collisionally ionized Fe XXV He$\alpha$ line at $\sim 6.7$ keV and Fe XXVI Ly$\alpha$ line at $\sim 6.9$ keV and their He$\alpha$ / Ly$\alpha$ line ratio \citep{Kurihara_2024} of $\sim 0.40$.

The Fe K$\alpha$ ($\sim 6.4$ keV) line was detected during the impulsive phase of the flare (Interval 0$-$2).
Then, to improve the fits, we added the \texttt{gauss} at 6.4 keV for Interval 0, 1, and 2.
The statistical significance and equivalent width of the Fe K$\alpha$ line at Interval 2 were $5.3 \sigma$ significance and $67^{+28}_{-20}$ eV, respectively.
Figure \ref{FigureE1} and Table \ref{TableE1} and \ref{TableE2} summarize the spectral fitting and its best-fitting parameters, respectively.
The Fe K$\alpha$ line fluxes in Table \ref{TableE2} does not include the nearby continuum emission.
Details of the assessment of the statistical significance is found in Appendix \ref{app:significance}

\begin{figure}
\begin{center}
\includegraphics[width=8.5cm]{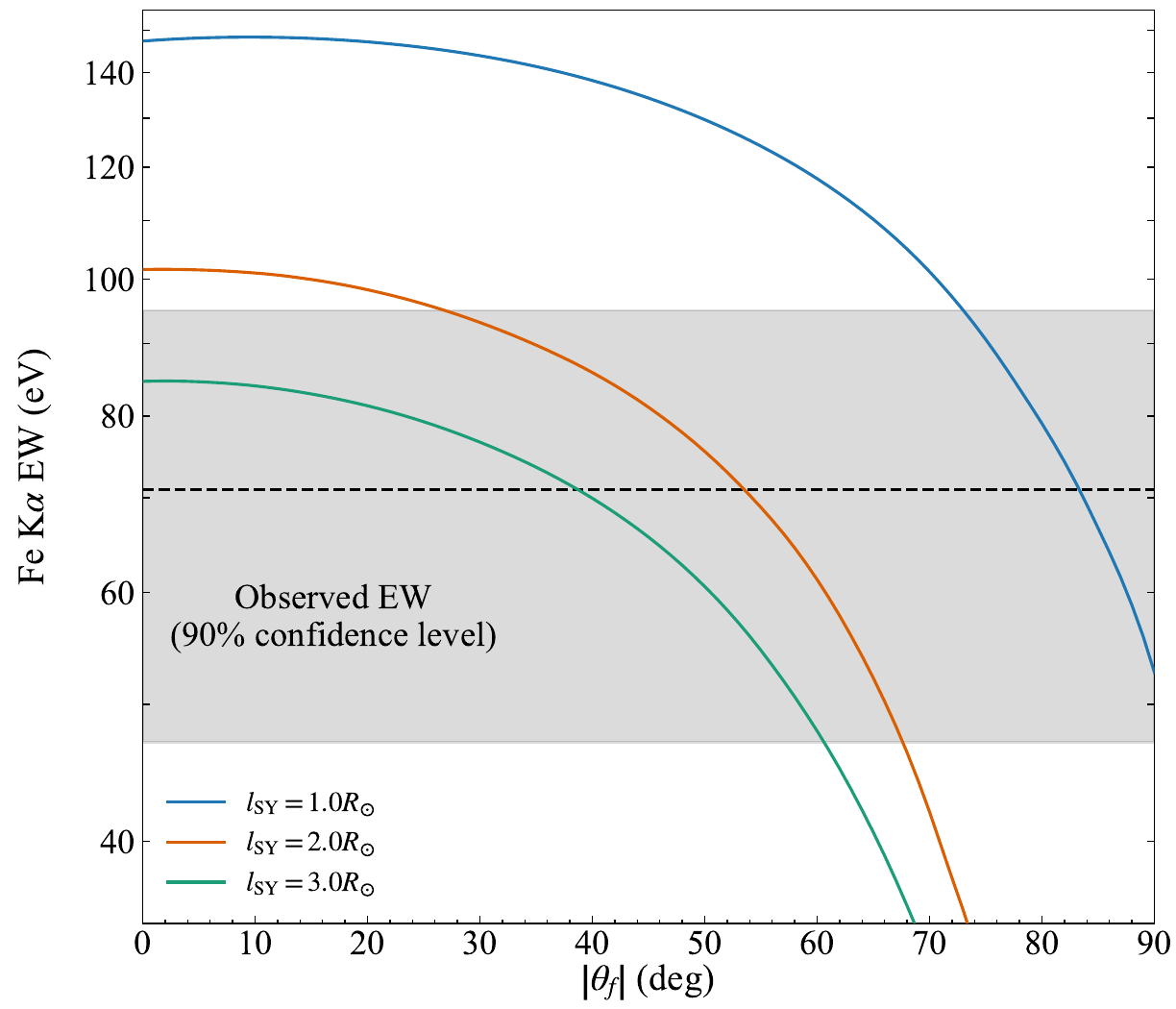}
\end{center}
\caption{
Comparison between the observed Fe K$\alpha$ line equivalent width and expectations as a function of $\theta_{f}$ obtained by \texttt{SKIRT}.  Blue, orange, and green lines correspond to $l_{\mathrm{SY}} = 1.0R_{\odot}$, $2.0R_{\odot}$, and $3.0R_{\odot}$ cases, respectively. The observed equivalent width with 90\% confidence level is shown as gray-shaded area. Note that $n_{\mathrm{p}}$, $\Delta r$, and $Z_{\mathrm{p}}$ are fixed to $10^{16}$ cm$^{-3}$, $0.01R_{\odot}$, and $Z_{\odot}$, respectively, in this calculation.
}
\label{Figure4}
\end{figure}

\section{Discussion and Conclusion} \label{sec:discussion_conclusion}
In the standard model of solar and stellar flares \citep[c.f.][]{Kowalski_2024}, FUV emission is emitted as a result of the initial abrupt heating of the transition region by non-thermal electrons \citep{Qiu_2021} while the X-ray emission occurs as the thermal plasma fills in the corona later \citep{Hirayama_1974}.  
As mentioned in Section \ref{subsec:analysis_result_lightcurve}, the time derivative of the soft X-ray emission during the impulsive phase well correlated with the FUV light curve (Figure \ref{Figure1}b). 
This kind of delay between a proxy for the non-thermal particles (e.g., FUV) and thermal (e.g., soft X-ray) emissions are consistent with so-called “Neupert effect”.

Figure \ref{Figure1}c shows the time evolution of the Fe K$\alpha$, He$\alpha$, and Ly$\alpha$ line luminosities.
These luminosities reached their maximum at the same time as that of the thermal soft X-ray continuum ($t\sim0.63$ day in Figure \ref{Figure1}). 
The time between the liberation of the electron by the X-ray photon and subsequent emission of the Fe K$\alpha$ line is $10^{-12} -10^{-9}$ sec \citep{Trabert_2023} and much less than the timescale of the light curve (Figure \ref{Figure1}c).
The coincidence of the Fe K$\alpha$ line peak with the thermal soft X-ray peak, not with the non-thermal FUV peak, strongly supports the thermal photoionization mechanism as the origin of the Fe K$\alpha$ line.

To test the validity of the photoionization scenario from the standpoint of the Fe K$\alpha$ line intensity and to demonstrate the potential of this line as a diagnostic of flare geometry, we performed 3D montecarlo radiative transfer calculations with \texttt{SKIRT v9.0} \citep{Vander_2023}.
Following \cite{Ercolano_2008}, we treated the intrinsic flare emission as a point source and assumed spherical thermal X-ray emission irradiates the stellar photosphere (see Figure \ref{Figure3} and Appendix \ref{app:calc_setup} for details of the simulation setup).
The X-ray flare peak spectrum in Interval 2 (Figure \ref{FigureE1}c) was adopted as the input spectral energy distribution (SED) of the source. 
The distance between the flare emission source and the irradiated gas was set to $r_{0} = l_{\mathrm{SY}}/2.0$, where $l_{\mathrm{SY}} = (0.2$–$3.9) R_{\odot}$ is the loop size estimated from the observed flare temperature and emission measure using the magnetic reconnection model of \cite{Shibata_2002} (see Appendix \ref{app:calc_setup} for details) and $R_{\odot}$ denotes the solar radius.
The thickness of the irradiated gas was fixed at $\Delta r = 0.01 R_{\odot}$, considering that the solar photospheric thickness is $\mathcal{O}(0.001 R_{\odot})$ and that the ratio of the scale height of UX Ari Aa to that of the Sun is $\mathcal{O}(10)$. 
The density of the irradiated gas (i.e., the photosphere) was set to $n_{\mathrm{p}} = 10^{16}$ cm$^{-3}$, a typical value for the solar photosphere.
Note that our calculation ignores absorption of X-rays in the chromosphere.
We varied the inclination angle between the observer and the flare emission source (Figure \ref{Figure3}) from $0^\circ \leq |\theta_{f}| \leq 90^\circ$ in steps of 1 deg.

Figure \ref{Figure4} compares the observed and modeled Fe K$\alpha$ line equivalent widths. 
We show the results of the calculations for $l_{\mathrm{SY}} = 1.0 R_{\odot}$, $2.0 R_{\odot}$, and $3.0 R_{\odot}$, which lie within the range $(0.2$–$3.9) R_{\odot}$. 
The observed equivalent width can be reproduced for any loop size, further supporting the photoionization scenario. 
The calculated equivalent width decreases with increasing $l_{\mathrm{SY}}$, reflecting the reduced fluorescence efficiency at larger loop heights and the dilution of the irradiating X-ray flux at the stellar surface \citep{Ercolano_2008, Inoue_2025}. 
The inclination angle required to match the observations varies with the loop size. 
For the $l_{\mathrm{SY}} = 1.0 R_{\odot}$ case, reproducing the observed equivalent width requires $75^\circ < |\theta_{f}| < 90^\circ$, corresponding to a flare latitude of $-60^\circ < 90^\circ - i_{r} + \theta_{f} < -45^\circ$ or $60^\circ < 180^\circ - (90^\circ - i_{r} + \theta_{f}) < 75^\circ$. 
In contrast, for the $l_{\mathrm{SY}} = 3.0 R_{\odot}$ case, the observed equivalent width requires $|\theta_{f}| < 60^\circ$, corresponding to a flare latitude of $-30^\circ < 90^\circ - i_{r} + \theta_{f} < 90^\circ$. 
These results indicate that the Fe K$\alpha$ line intensity can be used to constrain the parameter space of flare loop size and latitude. 
Tighter constraints on these parameters will require more precise measurements of the Fe K$\alpha$ equivalent width, which will be enabled by high-resolution spectroscopy with XRISM \citep{Tashiro_2025}.

In this study, we found clear evidence that photoionization is the dominant emission mechanism of the Fe K$\alpha$ line during a stellar flare. 
The coincidence of the Fe K$\alpha$ line peak with the thermal X-ray peak provides the strongest support for the photoionization scenario. 
Furthermore, our 3D radiative transfer calculations confirmed that the observed Fe K$\alpha$ line intensity is consistent with photoionization and demonstrated the potential of this line to constrain the flare loop size and latitude. 
These results suggest that future studies of stellar CMEs may incorporate the direction of motion as an additional diagnostic using the Fe K$\alpha$ line.

%TC:ignore

%% Please use the acknowledgment and contribution environments. This will 
%% be anonomyized when the "anonymous" style option is used. 
\begin{acknowledgments}
This research has made use of data and/or software provided by the High Energy Astrophysics Science Archive Research Center (HEASARC), which is a service of the Astrophysics Science Division at NASA/GSFC. NICER analysis software and data calibration were provided by the NASA NICER mission and the Astrophysics Explorers Program.
This research is supported by JSPS KAKENHI grant No. 24KJ1483 (S.I.) and a grant from the Hayakawa Satio Fund awarded by the Astronomical Society of Japan.
T.E. was supported by the JST, Japan grant number JPMJFR202O (Sohatsu).
Y.N. acknowledge funding support through  NASA NICER Cycle 6 Program 80NSSC24K1194 and NASA TESS Cycle 7 Program 80NSSC25K7906.
\end{acknowledgments}

\begin{contribution}
%%This section gives authors the space to recognize author contributions. The text inside this environment is NOT counted towards the total word quanta. At a minimum, manuscripts are expected to include this text:

S.I. conducted the data analyses and simulations and wrote the draft of the manuscript. W.B.I. and T.K. led the campaign observations and contributed to the data analysis. T.E., Y.N., H.U., K.H., and S.T. contributed to the data interpretation, scientific discussions and text writing. A.Y., F.T., G.M., K.Y., Z.A., and K.G. contributed to the satellite operation and observations.

%% But authors are expected to provide more specific details, e.g. 
%%
%%SC was responsible for writing and submitting the manuscript.
%%WWM came up with the initial research concept and edited the manuscript.
%%OTS obtained the funding and edited the manuscript.
%%EBF provided the formal analysis and validation. He also edited the manuscript.
%%GEH Supervised the undergraduates, wrote the software and administers the project github and Zenodo repositories.
%%
%% Authors can use the Contributor Role Taxonomy (CRediT) at
%% for ideas on how write a good statement tailored to their needs.

\end{contribution}

\facilities{Hisaki, NICER \citep{2016SPIE.9905E..1HG}}

\software{astropy \citep{astropy:2013, astropy:2018}, HEAsoft \citep{heasoft_2014}, Xspec \citep{1996ASPC..101...17A}, \texttt{SKIRT v9.0} \citep{Vander_2023}}


\appendix
\restartappendixnumbering 
\section{Light curves of each FUV line}\label{app:FUV_lines}
\begin{figure}
\begin{center}
\includegraphics[width=8cm]{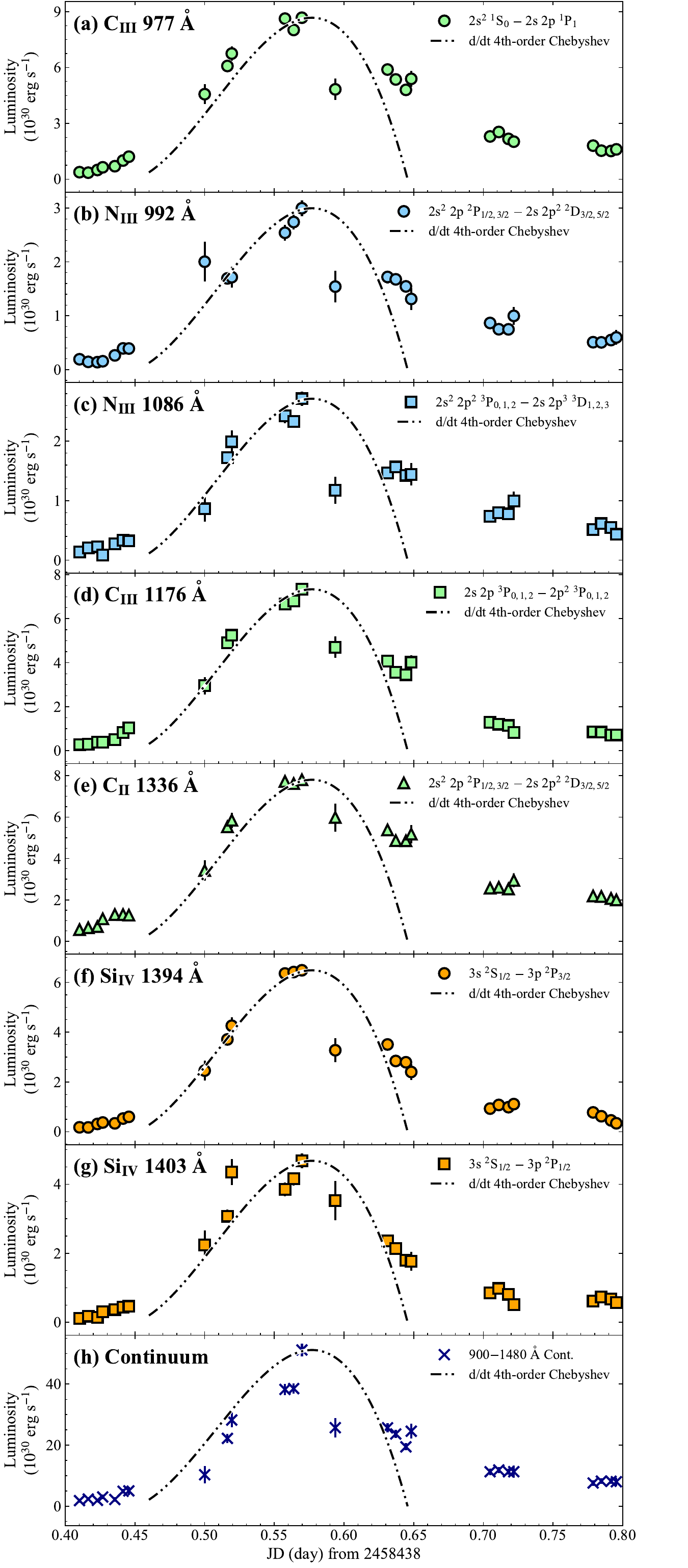}
\end{center}
\caption{
Time variation of the luminosity of (a-g) FUV emission lines and (h) continuum in 900$-$1480 \AA. Black dash-dot-dot lines show the time derivative of the 4-th order Chebyshev function fitted for the impulsive phase of the soft X-ray light curve (Figure \ref{Figure1}). The derivative curves of the Chebyshev function are normalized to the peak luminosity of each line.
}
\label{FigureE0}
\end{figure}

We display the time variation of the luminosity of each FUV line and continuum in 900$-$1480 \AA with the time derivative of soft X-ray light curve in Figure \ref{FigureE0}.

\section{Best-fit spectral parameters }\label{app:best-fit}
We provide the best-fit parameters of the X-ray spectra during the quiescence and each each Interval during the flare in Table \ref{TableE1} and \ref{TableE2}, respectively.

\input{tabE1}
\input{tabE2}

\section{Possibility of Fe L$\alpha$ fluorescence emission}\label{app:FeLa}

\begin{figure}
\begin{center}
\includegraphics[width=0.48\linewidth,clip]{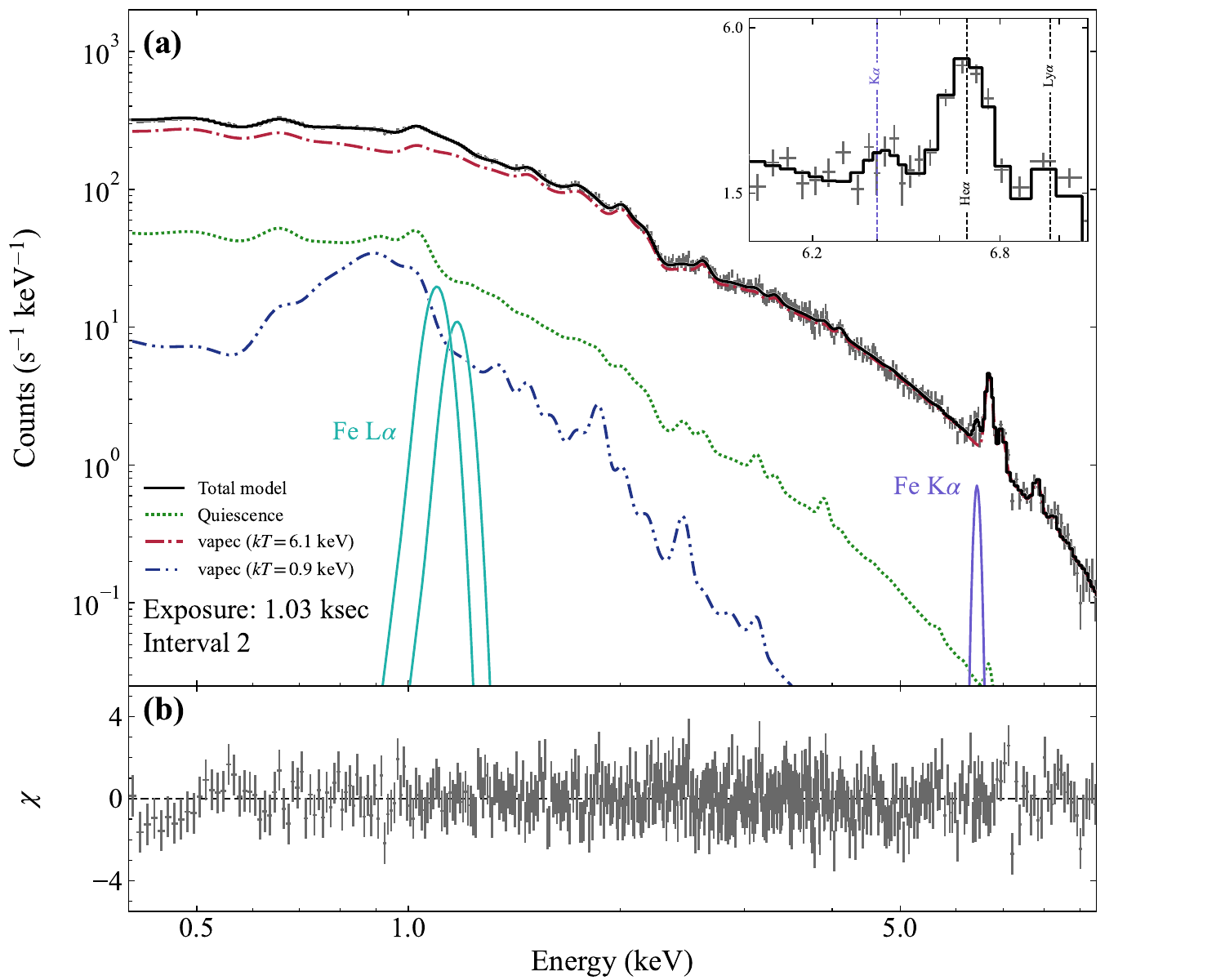}
\end{center}
\caption{Spectral fitting of the Interval 2 spectrum with the two-temperature CIE and gaussians for the Fe L$\alpha$ and K$\alpha$ lines. (a) The total best-fit model (solid black line) is composed of high temperature ($kT = 6.1$ keV; red dash-dot line) and low temperature ($kT = 0.9$ keV; blue dash-dot-dot line) CIE (\texttt{vapec}) components, fixed quiescence component (green dotted line) and Gaussian emission lines at $\sim 1.1$ keV (turquoise solid lines) and $\sim 6.4$ keV (purple solid line). The right upper inset panel
shows the enlarged spectrum around the Fe K$\alpha$ at 6.4 keV, Fe XXV He$\alpha$ at 6.7 keV, and Fe XXVI Ly$\alpha$ at 6.9 keV. (b) The residuals between the data and model shown in panel a. The errorbars indicate 1$\sigma$ uncertainties.
}
\label{Figure2_old}
\end{figure}

We also considered the possibility that Fe L$\alpha$ fluorescence emission from the photosphere \citep{Drake_1999} contributes to the residuals at 1.0–1.2 keV in the Interval 2 spectrum when using the two-temperature CIE model. Adding two Gaussian components for the Fe L$\alpha$ line yielded best-fit line energies of $1.09^{+0.02}_{-0.02}$ and $1.17^{+0.03}_{-0.03}$ keV (Figure \ref{Figure2_old}). However, the inferred ionization states of these Fe L$\alpha$ lines, based on their center energies, are much higher than that of the Fe K$\alpha$ line \citep{Kallman_1995}. We therefore refrain from attributing these features to photoionized plasma emission and instead adopt the three-temperature CIE model as described in Section \ref{subsec:analysis_result_spectrum}.

\section{Assessment of the statistical significance of the Fe K$\alpha$ line}\label{app:significance}

\begin{figure}
\begin{center}
\includegraphics[width=0.47\linewidth,clip]{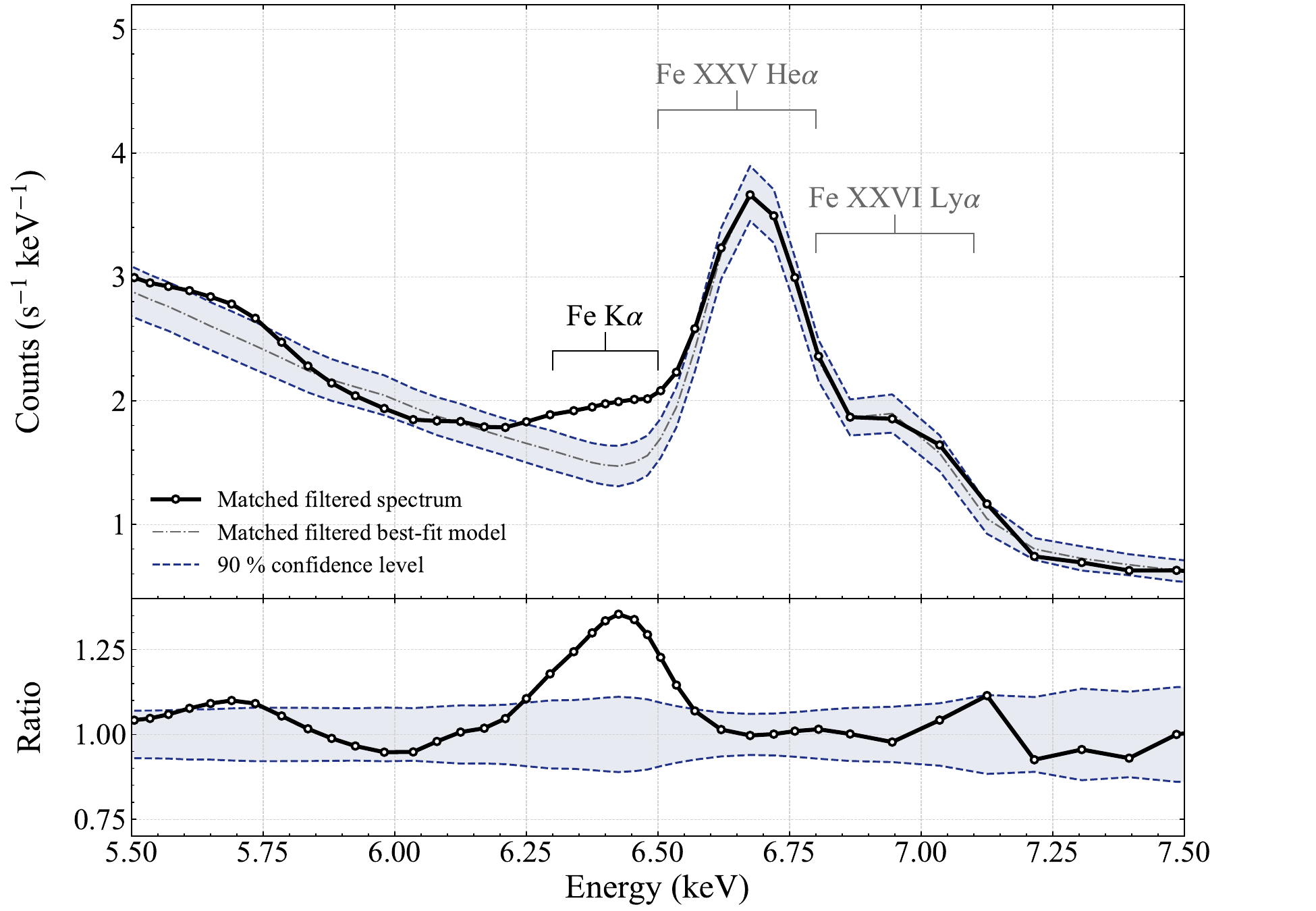}
\end{center}
\caption{Matched-filtering line search results of the Interval 2 spectrum. The solid black line and gray dashdot line indicate the matched-filtered observed spectrum and best-fitting model for the energy band except for the Fe  K$\alpha$ line energy. The blue shaded area shows the 90 \% confidence level of the best-fit model (\texttt{apec}), respectively. 
(b) The ratio of the matched filtered spectrum and statistical fluctuations of the continuum model to the best-fit model.
}
\label{FigureE2}
\end{figure}

We investigated the significance of the Fe K$\alpha$ lines in the Interval 2 spectrum using the matched-filtering line search method \citep{Rutledge_2003, Hurkett_2008, Miyazaki_2016, Inoue_2025}.
The procedure of our investigation is as follows:
\begin{enumerate}
    \item We fitted the 5$-$8 keV Interval 2 spectrum with the simple CIE model (\texttt{apec}) ignoring the Fe K$\alpha$ bands (i.e., 5$-$8 keV except for 6.20$-$6.55 keV). The CIE model includes the collisionally ionized Fe XXV He$\alpha$ line at $\sim 6.7$ keV and Fe XXVI Ly$\alpha$ line at $\sim 6.9$ keV. Then, we generated $10^{4}$ simulated spectra with the \texttt{fakeit} command in \texttt{Xspec} assuming the parameters of the best-fit CIE model without the Fe K$\alpha$ line.
    \item We processed the $10^{4}$ simulated spectra with the matched filter (Equation 1 of \cite{Miyazaki_2016}). In this process, we calculated the full width at half-maximum (FWHM) of NICER from the Response Matrix File (\texttt{rmf}) file, which was also used to fit the observed Interval 2 spectrum.
    \item We also applied the same matched filter to the observed Interval 2 spectrum (Figure \ref{FigureE1}) to maximize the signal-to-noise ratio. 
    \item We investigated the statistical distribution of the count rates of the matched filtered fake spectra for each energy bin and calculated the significance of the Fe K$\alpha$ line.
\end{enumerate}

Extended Data Figure \ref{FigureE2} shows the result of the significance study.
The count rates around the line center of the Fe K$\alpha$ line comes above the 90 \% confidence level of the best-fit model without this line.
As a result, the significance of the Fe K$\alpha$ line was estimated to be at $5.3 \sigma$ level.

\section{Radiative transfer calculation setup}\label{app:calc_setup}

For the radiative transfer calculation using \texttt{SKIRT}, we need to set three major assumptions; (1) the distance between the irradiated gas and emission source, (2) the thickness of the irradiated gas, and (3) the iron abundance and density of the photosphere.
\begin{enumerate}
    \item We estimate the distance between the irradiated gas of the stellar surface and X-ray emission source of the flare as 
\begin{equation}
   % F_{\mathrm{X}}^{\mathrm{photo}} \sim \frac{L_{\mathrm{X}}^{\mathrm{flare}}}{4 \pi (l_{\mathrm{SY}}/2)^{2}}, \label{eq:photosphere_flux}
   r_{0} = l_{\mathrm{SY}}/2,
\end{equation}
where $l_{\mathrm{SY}}$ is the flare loop size calculated by the magnetic reconnection model equations in \cite{Shibata_2002},
\begin{equation*} 
l_{\mathrm{SY}} = 10^{9} \left( \frac{EM_{\mathrm{peak}}}{10^{48} \: \mathrm{cm}^{-3}} \right)^{3/5} \nonumber \times \left( \frac{\mathnormal{n_{0}}}{10^{9} \: \mathrm{cm}^{-3}} \right)^{-2/5} \left( \frac{\mathnormal{T_{\mathrm{peak}}}}{10^{7} \: \mathrm{K}} \right)^{-8/5}
\: \mathrm{cm}. \label{eq:Shibata_Yokoyama_L}
\end{equation*}
Here, $EM_{\mathrm{peak}}$ is the volume emission measure at the flare peak, $T_{\mathrm{peak}}$ is the peak electron temperature, and $n_{0}$ is the preflare coronal density.
We substituted the total emission measure ($1.2\times10^{55}$ cm$^{-3}$) and emission measure weighted temperature (6.7 keV) of the Interval 2 (Table \ref{TableE2}) for $EM_{\mathrm{peak}}$ and $T_{\mathrm{peak}}$ in Equation \ref{eq:Shibata_Yokoyama_L}.
We assumed the preflare coronal density to be $n_{0} = 10^{10-13} \: \mathrm{cm^{-3}}$ \citep{Aschwanden_1997, Gudel_2004, Reale_2007, Sasaki_2021} in this calculation.
As a result, the $l_{\mathrm{SY}}$ and $r_{0}$ are $(0.2-3.9) R_{\odot}$ and $(0.1-2.0) R_{\odot}$, respectively.

    \item We estimated the thickness of the photosphere of UX Ari Aa by calculating the scale height and used the value as the thickness of the irradiated gas. The ratio of the scale height of UX Ari Aa to that of the Sun is 
\begin{equation} 
    \label{eq:H_ratio}
    \frac{H_{U}}{H_{\odot}} = \frac{T_{U}}{T_{\odot}} \times \frac{g_{\odot}}{g_{U}} \sim 2 \times 10^{1},
\end{equation}
where $T_{U} = 4560$ K and $T_{\odot} = 5780$ K are the effective temperature and $g_{U} = 1.14 \times 10^{3}$ cm s$^{-2}$ and $g_{\odot} = 2.74 \times 10^{4}$ cm s$^{-2}$ are the surface gravity of UX Ari \citep{Hummel_2017} and the Sun, respectively.
We assumed that the mass and radius of UX Ari Aa are $1.30 M_{\odot}$ and $5.6 R_{\odot}$, respectively, where $M_{\odot}$ and $R_{\odot}$ are the mass and radius of the Sun, respectively.  
Considering the thickness of the solar photosphere of $\mathcal{O} (10^{-3} \times R_{\odot})$ and the scale height ratio between the Sun and UX Ari Aa (Equation \ref{eq:H_ratio}), we estimated the thickness of the photosphere of UX Ari Aa to be $0.01 \times R_{\odot}$.
We thus set the thickness of the irradiated gas $\Delta r = 0.01 \times R_{\odot}$.

   \item The abundances of all elements in the irradiated gas ($Z_{\mathrm{p}}$) were set to solar photospheric values $Z_{\odot}$ \citep{Anders_1989}, as no measurements are available for UX Ari Aa. The gas density was set to $n_{\mathrm{p}} = 10^{16}$ cm$^{-3}$, a typical solar photospheric value \citep{Bommier_2020}. This corresponds to a vertical column density of
\begin{equation}
    N_{\mathrm{p}} = n_{\mathrm{p}} \Delta r \sim 7 \times 10^{24} \; \mathrm{cm^{-2}}.
\end{equation}
Thus, the photospheric thickness $\Delta r$ is thick enough for X-rays to emit fluoresce Fe K$\alpha$ line in this simulation.
\end{enumerate}

Under these assumptions, we performed radiative transfer calculations for $l_{\mathrm{SY}} = 1.0 R_{\odot}$, $2.0 R_{\odot}$, and $3.0 R_{\odot}$ with the inclination angles between the observer and the flare emission source in the range $0^\circ \leq |\theta_{f}| \leq 90^\circ$ using \texttt{SKIRT v9.0} \citep{Vander_2023}.  
\texttt{SKIRT} is a 3D radiative transfer code that incorporates Compton scattering by free electrons, photoabsorption and fluorescence by cold atomic gas, scattering by bound electrons, and dust extinction \citep{Vander_2023}.  
We adopted the X-ray flare peak spectrum from Interval 2 (Figure \ref{FigureE1}c) and its 5–15 keV luminosity of $9.8 \times 10^{31}$ erg s$^{-1}$ as the input SED and luminosity.  
The numbers of seed pseudo-photons required to obtain sufficient statistics were $10^{10}$ for the $l_{\mathrm{SY}} = 1.0 R_{\odot}$ case and $10^{9}$ for the $l_{\mathrm{SY}} = 2.0 R_{\odot}$ and $3.0 R_{\odot}$ cases.  
When the distance between the source and the irradiated gas decreases, photons penetrate deeper into the gas and undergo more interactions, thus requiring larger input photon statistics.
The equivalent width of the Fe K$\alpha$ line was obtained from the output spectra after convolution with the NICER energy resolution of 140 eV at 6 keV.  
Note that \texttt{SKIRT v9.0} employs the solar photospheric abundance table of \cite{Anders_1989}, in which the iron abundance is $\sim$1.5 times higher than in other tables \citep{Grevesse_1998, Wilms_2000, Asplund_2009}. Consequently, the results carry an uncertainty of approximately this factor.

%% For this sample we use BibTeX plus aasjournalv7.bst to generate the

\bibliography{sample701}{}
\bibliographystyle{aasjournalv7}



%TC:endignore
\end{document}

%% file: tabE1.tex
\renewcommand{\arraystretch}{0.9}
\begin{table}
\centering
\caption{Best-fit parameters of the quiescence (ObsID: 1100380107) spectrum with three temperature collisionally-ionized models. The error ranges correspond to the 90\% confidence level. Values without errors are fixed. The norm of \texttt{vapec} means $10^{-14} (4 \pi)^{-1} [D_{\mathrm{A}} (1+z)]^{-2}] \int n_{\mathrm{e}} n_{\mathrm{H}} dV$, where $D_{\mathrm{A}}$ is the angular diameter distance to the source, $z$ is the redshift, $n_{\mathrm{e}}$ and $n_{\mathrm{H}}$ are the electron and hydrogen densities, and $dV$ is the volume element. $F_{\mathrm{X}}$ is the 0.3$-$4 keV flux.}
\label{TableE1}
\vspace{0.3cm}
\begin{tabular}{ccc}
\hline 
\multicolumn{2}{c}{Exposure (ks)} &  $4.00$  \\ \hline
\multicolumn{2}{c}{\texttt{tbabs}} &  \\
\multicolumn{2}{c}{$N_{\mathrm{H}}$ ($10^{19}$ $\mathrm{cm^{-2}}$)} & $5.90_{-5.39}^{+5.88}$ \\ \hline
\multicolumn{2}{c}{\texttt{vapec} (Low Temp.)} &  \\
\multirow{2}{*}{Temperature} & $kT$ (keV) & $0.51_{-0.04}^{+0.05}$ \\
& $T$ (MK) & $5.86_{-0.47}^{+0.61}$ \\
\multicolumn{2}{c}{norm ($10^{-2}$)} & $0.99_{-0.36}^{+0.48}$ \\ \hline
\multicolumn{2}{c}{\texttt{vapec} (Med Temp.)} &  \\
\multirow{2}{*}{Temperature} & $kT$ (keV) & $0.91_{-0.06}^{+0.06}$ \\
& $T$ (MK) & $10.54_{-0.75}^{+0.68}$ \\
\multicolumn{2}{c}{norm ($10^{-2}$)} & $1.72_{-0.38}^{+0.46}$ \\ \hline
\multicolumn{2}{c}{\texttt{vapec} (High Temp.)} &  \\
\multirow{2}{*}{Temperature} & $kT$ (keV) & $2.42_{-0.13}^{+0.20}$ \\
& $T$ (MK) & $28.11_{-1.56}^{+2.34}$ \\
\multicolumn{2}{c}{norm ($10^{-2}$)} & $3.50_{-0.31}^{+0.25}$ \\ \hline
\multirow{11}{*}{$Z$ ($Z_{\odot}$)} 
& He & $1.00$ \\
& C, N, O & $0.36_{-0.06}^{+0.07}$  \\
& Ne & $1.11_{-0.22}^{+0.26}$  \\
& Mg & $0.12_{-0.06}^{+0.07}$  \\
& Al & $1.00$ \\
& Si & $0.18_{-0.04}^{+0.05}$  \\
& S & $0.34_{-0.08}^{+0.08}$  \\
& Ar & $1.00$ \\
& Ca & $1.00$ \\
& Fe & $0.06_{-0.01}^{+0.02}$  \\
& Ni & $1.00$ \\
\hline
\multicolumn{2}{c}{$F_{\mathrm{X}}$ ($10^{-11}$ erg cm$^{-2}$ s$^{-1}$)} & $5.91^{+0.02}_{-0.06}$ \\ \hline
\multicolumn{2}{c}{$\chi^{2}$ ($d.o.f$)} & $265.27$ ($233$) \\ \hline
\end{tabular}
\end{table}

%% file: tabE2.tex
\renewcommand{\arraystretch}{0.9}
\begin{table}
\centering
\caption{Best-fit parameters of the flare (ObsID: 1100380108) spectra with the sum of the fixed quiescence component, two temperature collisionally-ionized models, and gaussian for the Fe K$\alpha$ line. \label{TableE2}}
\vspace{0.3cm}
\begin{tabular}{cccccc} \hline
\multicolumn{2}{c}{} & Interval 0 & Interval 1 & Interval 2 & Interval 3  \\ \hline 
\multicolumn{2}{c}{Exposure (ks)} & $1.13$ & $1.17$ & $1.03$ & $0.62$   \\ \hline
\multicolumn{2}{c}{\texttt{tbabs}} & & & &   \\
\multicolumn{2}{c}{$N_{\mathrm{H}}$ ($10^{19}$ $\mathrm{cm^{-2}}$)} 
& $5.90$ & $5.90$ & $5.90$ & $5.90$  
\\ \hline
\multicolumn{2}{c}{\texttt{vapec} (High Temp.)} & & & &   \\
\multirow{2}{*}{Temperature} & $kT$ (keV) 
& $4.30_{-0.36}^{+0.43}$ & $8.63_{-0.66}^{+0.93}$ & $7.21_{-0.38}^{+7.70}$ & $3.94_{-0.12}^{+0.12}$ 
\\
& $T$ (MK) 
& $49.84_{-4.16}^{+5.0}$ & $100.11_{-7.66}^{+10.79}$ & $83.64_{-4.41}^{+89.32}$ & $45.75_{-1.4}^{+1.4}$  
\\
\multicolumn{2}{c}{norm ($10^{-2}$)} 
& $1.98_{-0.12}^{+0.12}$ & $16.66_{-1.60}^{+1.05}$ & $36.64_{-22.42}^{+1.64}$ & $14.79_{-0.4}^{+0.4}$  
\\ \hline
\multicolumn{2}{c}{\texttt{vapec} (Mid Temp.)} & & & & \\
\multirow{2}{*}{Temperature} & $kT$ (keV) 
& --- & $1.97_{-0.51}^{+0.85}$ & $1.92_{-0.33}^{+2.65}$ & ---  
\\
& $T$ (MK) 
& --- & $22.85_{-5.92}^{+9.86}$ & $22.27_{-3.83}^{+30.74}$ & ---
\\
\multicolumn{2}{c}{norm ($10^{-2}$)} 
& --- & $1.66_{-0.76}^{+1.47}$ & $3.59_{-1.32}^{+19.68}$ & --- 
\\ \hline
\multicolumn{2}{c}{\texttt{vapec} (Low Temp.)} & & & & \\
\multirow{2}{*}{Temperature} & $kT$ (keV) 
& $0.79_{-0.04}^{+0.04}$ & $0.78_{-0.04}^{+0.03}$ & $0.80_{-0.06}^{+0.06}$ & $0.84_{-0.04}^{+0.04}$  
\\
& $T$ (MK) 
& $9.22_{-0.46}^{+0.45}$ & $9.05_{-0.46}^{+0.35}$ & $9.28_{-0.70}^{+0.70}$ & $9.72_{-0.51}^{+0.52}$   
\\
\multicolumn{2}{c}{norm ($10^{-2}$)} 
& $0.27_{-0.06}^{+0.08}$ & $0.60_{-0.06}^{+0.07}$ & $0.49_{-0.07}^{+0.07}$ & $0.94_{-0.12}^{+0.13}$  
\\ \hline
\multirow{11}{*}{$Z$ ($Z_{\odot}$)} 
& He & $1.00$  & $1.00$  & $1.00$  & $1.00$  \\
& C, N, O & $1.31_{-0.51}^{+0.64}$  & $1.33_{-0.28}^{+0.30}$  & $1.09_{-0.20}^{+0.21}$  & $0.99_{-0.21}^{+0.22}$  \\
& Ne & $1.00$  & $2.44_{-1.04}^{+0.89}$  & $2.40_{-0.71}^{+1.39}$  & $3.41_{-0.57}^{+0.58}$  \\
& Mg & $1.00$  & $1.22_{-0.37}^{+0.39}$  & $1.37_{-0.28}^{+0.29}$  & $1.00$  \\
& Al & $1.00$  & $2.53_{-2.53}^{+3.72}$  & $7.03_{-2.72}^{+2.78}$  & $1.00$  \\
& Si & $1.00$  & $1.53_{-0.27}^{+0.29}$  & $1.30_{-0.18}^{+0.19}$  & $0.82_{-0.21}^{+0.21}$  \\
& S & $1.00$  & $1.27_{-0.43}^{+0.45}$  & $1.32_{-0.27}^{+0.29}$  & $0.39_{-0.31}^{+0.31}$  \\
& Ar & $5.61_{-2.86}^{+2.97}$  & $2.78_{-1.34}^{+1.40}$  & $1.84_{-0.78}^{+0.80}$  & $1.00$  \\
& Ca & $1.00$  & $2.71_{-1.56}^{+1.56}$  & $2.37_{-0.90}^{+0.90}$  & $1.00$  \\
& Fe & $0.56_{-0.14}^{+0.16}$  & $0.71_{-0.08}^{+0.08}$  & $0.70_{-0.05}^{+0.11}$  & $0.46_{-0.05}^{+0.05}$  \\
& Ni & $1.00$  & $1.39_{-1.07}^{+1.13}$  & $2.21_{-0.67}^{+0.70}$  & $1.00$  \\
\hline
\multicolumn{2}{c}{\texttt{gauss} (Fe K$\alpha$)} & & & & \\
\multicolumn{2}{c}{$E_{l}$ (keV)} & $6.38_{-0.06}^{+0.07}$ & $6.44_{-0.04}^{+0.04}$ & $6.42_{-0.03}^{+0.04}$ & --- \\
\multicolumn{2}{c}{norm ($10^{-4}$ Photons cm$^{-2}$ s$^{-1}$)}  & $0.48_{-0.33}^{+0.30}$ & $1.72_{-0.75}^{+0.75}$ & $3.35_{-1.11}^{+1.11}$ & --- \\ 
\multicolumn{2}{c}{EW (eV)}  & $192_{-130}^{+123}$ & $73^{+29}_{-29}$ & $67^{+28}_{-20}$ & --- \\ \hline
\multicolumn{2}{c}{$F_{\mathrm{X}}$ ($10^{-10}$ erg cm$^{-2}$ s$^{-1}$)} & $0.91^{+0.01}_{-0.01}$ & $3.11^{+0.01}_{-0.02}$ & $5.88^{+0.02}_{-0.02}$ & $2.65^{+0.01}_{-0.02}$ \\ \hline
\multicolumn{2}{c}{$\chi^{2}$ ($d.o.f$)} & $230.06$ ($218$) & $413.17$ ($439$) & $418.99$ ($469$) & $259.54$ ($264$) \\ \hline
% \tablecomments{The error ranges correspond to $90\%$ confidence level. Values without errors mean that they are fixed.}
\end{tabular}
\end{table}